# Response to "Comment on 'Electric Multipole Moments for Some First-Row Diatomic Hydride Molecules' [Commun. Theor. Phys. 38 (2002) 256]"

Metin Orbay[*] and Telhat Ozdogan

Department of Physics, Faculty of Education, Ondokuz Mayis University, Amasya, Turkey

**Abstract** The comment of Guseinov is irrelevant and also unjust. In contrast to his comment, we show that the obtained electric multipole moment values for some first-row diatomic molecules are original and better than his values (I.I. Guseinov, E. Akin, and A.M. Rzaeva, J. Mol. Struct. (Theochem) **453** (1998) 163) with respect to Hartree-Fock values. Moreover, it must be noted that all the formulas are cited in our paper (M. Orbay and T. Ozdogan, Commun. Theor. Phys. (Beijing, China) **35** (2001) 585) and corrigendum (M. Orbay and T. Ozdogan, Commun. Theor. Phys. (Beijing, China) 37 (2002) 768).

**PACS** numbers: 31.20.Di, 31.20.Ej, 31.90.+s

The comment of Guseinov[1] includes some claims that we consider as irrelevant and also unjust. With respect to his comments, his claims will be replied in the following:

*Comment 1* He claims that the formulas given by Eqs. (1)-(6) in our paper[2] are copied directly from his paper[3].

*Response to comment 1*

- As stated in our paper,[2] equations (1) and (2) are originally introduced in Ref. [4] (cited as Ref. [17] in our paper [2]), not in his paper[3].

- Equations (3) and (5) are cited in corrigendum [5], which can also be found elsewhere (see Refs. [6] and [7]).

- As is well known in literature, equation (4) is called Linear Combination of Atomic Orbitals (LCAO), in troduced by Roothaan[8] (cited as Ref. [19] in our paper [2]), not by Guseinov.

- From the fundamental postulates of quantum mechanics, expectation values of any physical quantity is expressed as in Eq. (6) (see Ref. [9]), not introduced by Guseinov!

*Comment 2* He claims that the obtained electric multipole moment values for some first-row diatomic hydride molecules are not original.

*Response to comment 2* As stated in our paper[2], the expectation values of electric multipole moments are expressed through the electric multipole moment integrals and linear combination coefficients. That is, the efficiency of electric multipole moments strongly depends on accurate and speed calculation of electric multipole moment integrals and linear combination coefficients. In this respect, our study[2] has the following advantages over his study[3]:

- The efficiency of electric multipole moment integrals depends on accurate and speed calculation of overlap integrals (see Eq. (7) of Ref. [10]). In the calculation of electric multipole moment integrals, Guseinov[3] used the formula for overlap integrals presented in Ref. [11], which we think the efficiency may decrease especially in higher values of quantum numbers and internuclear distances due to the mathematical auxiliary functions appearing in structure of overlap integrals. But in our study, we used the recurrence relation for overlap integrals presented in Ref. [12], which do not injure from possible instability problems. Also, it can be seen easily from the Tables of Refs. [11] and [12] that the algorithm[12] used in our study[2] is more speed than the algorithm[11] used in Ref. [3]. It should be stressed that the used algorithm for the evaluation of multicenter electric multipole moment integrals are analyzed for wide changes of molecular parameters, and more advantages are obtained[13,14].

- In our study[2], the linear combination coefficients appearing in the evaluation of electric multipole moments of molecules LiH, BH and FH are calculated by constructing computer programs via the recent developments in the calculation of multicenter molecular integrals over Slater-type orbitals, under research fund with our group[15,16]. On the other hand, in his study[3] he used very old linear combination coefficients in Ref. [17], which may strongly effect the efficiency of electric multipole moment values.

---

[*]Correspondence to: telhatoz@omu.edu.tr

- As is known it is not possible to have empirical values of higher electric multipole moments for molecules, such as octupole and hexadecapole moments. Therefore, the theoretical determination of higher electric multipole moments attracts considerable attention[18]. While his study is limited to the calculation of octupole moments, our study[2] contains also hexadecapole moments.

- As is well known, owing to its modest size, Hydrogen Floride (HF) has served as a benchmark molecule for *ab initio* quantum chemical methods for over 40 years (see Ref. [19] and references quoted therein). In our study [2], the electric multipole moments, total energies, orbital energies and linear combination coefficients of molecular orbitals for molecule HF are presented. But his study [3] does not contain any calculations on this molecule.

*Comment 3* In his note, he claims that the authors made an article [20] by changing the summation indices of his formula.

*Response to comment 3* As can be easily seen from the title of our paper in Ref. [20] that our work includes the beginning step for the calculation of multicenter integrals over STOs with noninteger principal quantum numbers. And also, it is known that there was no study of Guseinov on the calculation of multicenter integrals over STOs with noninteger principal quantum numbers, up to the publication date of our work[20]. Moreover, the comment made by Guseinov[21] is replied and the advantages over his algorithm are critically analyzed in Ref. [22]*

---

*The response to comment in Ref. [21] should be addressed.